\documentclass[aps,prb,final,twocolumn,superscriptaddress]{revtex4-1}

\usepackage{graphicx}
\usepackage{epsfig}
\usepackage{natbib}

\newcommand{\CrNbS}{$\mathrm{CrNb}_3\mathrm{S}_6$}

\begin{document}


\title{Nucleation, instability, and discontinuous phase transitions in 
monoaxial helimagnets with oblique fields}



\author{Victor Laliena}
\email[]{laliena@unizar.es}
\affiliation{Instituto de Ciencia de Materiales de Arag\'on 
(CSIC -- University of Zaragoza), C/Pedro Cerbuna 12, 50009 Zaragoza, Spain}

\author{Javier Campo}
\email[]{javier.campo@csic.es}
\affiliation{Instituto de Ciencia de Materiales de Arag\'on 
(CSIC -- University of Zaragoza), C/Pedro Cerbuna 12, 50009 Zaragoza, Spain}
\affiliation{Centre for Chiral Science, Hiroshima University, Higashi-Hiroshima, 
Hiroshima 739-8526, Japan}

\author{Yusuke Kousaka}
\affiliation{Department of Chemistry, Faculty of Science, Hiroshima University, 
Higashi-Hiroshima, Hiroshima 739-8526, Japan}
\affiliation{Centre for Chiral Science, Hiroshima University, Higashi-Hiroshima, 
Hiroshima 739-8526, Japan}

\begin{abstract}
The phase diagram of the monoaxial chiral helimagnet as a function of temperature ($T$) and
magnetic field with components perpendicular ($H_x$) and parallel ($H_z$) to the chiral axis is 
theoretically studied via the variational mean field approach in the continuum limit. 
A phase transition surface in the three dimensional thermodynamic space separates a 
chiral spatially modulated phase from a homogeneous forced ferromagnetic phase.
The phase boundary is divided into three parts: two surfaces of
second order transitions of \textit{instability} and \textit{nucleation} type, in De Gennes 
terminology, 
are separated by a surface of first order transitions. Two lines of tricritical points
separate the first order surface from the second order surfaces. 
The divergence of the period of the modulated state on the \textit{nucleation}
transition surface has the logarithmic behavior typical of a chiral soliton lattice.
The specific heat diverges on the \textit{nucleation} surface 
as a power law with logarithmic corrections, while it shows a finite discontinuity 
on the other two surfaces.
The soliton density curves are described by a universal function of $H_x$ 
if the values of $T$ and $H_z$ determine a transition point lying on the \textit{nucleation}
surface; otherwise, they are not universal.
\end{abstract}

\pacs{111222-k}

\maketitle


\section{Introduction \label{sec:intro}}

Chiral magnets are currently the subject of intense investigations both because of their 
practical applications in technology and their interesing properties from the point of view of
fundamental science.
The applications exploit the charge and spin transport properties of a chiral magnet,
which are strongly affected by the magnetic structure and thus can be controlled by the
application of suitable magnetic fields \cite{Wolf01,Zutic04}.
In addition, due to its topological nature, the magnetic structure of a chiral magnet is 
protected against continuous deformations to homogeneous magnetic states, as ferromagnetic 
states. The chiral state can only turn into a homogeneous state through phase transitions
that take place at definite points of the phase diagram.
This robustness makes chiral magnets excelent candidates as the main components of spintronic
devices \cite{Fert13} and, for instance, they are specially suitable as
the main components of information storage devices \cite{Romming13}.

Besides the applications, chiral magnets are very interesting objects
from a fundamental point of view, as chiral symmetry and its breaking and restoration 
are ubiquitous phenomena appearing virtually in any domain of science, from elementary particle 
physics to astrophysics, and including chemistry, biology, and geology \cite{Wagniere07}.

In the monoaxial helimagnet \cite{Kishine15}, the competition between the ferromagnetic (FM) and 
Dzyaloshinskii--Moriya \cite{Dzyal58,Moriya60} (DM) interactions at low $T$ give rise to a 
spatially modulated chiral 
magnetic structure that, in absence of an applied field, has the form of an helix propagating 
with period $L_0$ along the chiral axis, which is called here the DM axis. 
At a certain ordering temperature, $T_0$, a magnetic transition to a paramagnetic (PM) phase takes place.
The period $L_0$ is independent of $T$, but the local magnetic moment 
decreases with $T$, and the transition to the PM state takes place at the temperature where it vanishes.
The nature of the transition at $T_0$ is not fully understood and considerable effort 
is being devoted to clarify this interesting question
\cite{Tsuruta16,Bornstein15,Ghimire13,Chapman14,Shinozaki15,Laliena16b,Nishikawa16}.

At temperatures lower than $T_0$, application of a magnetic field perpendicular to the DM axis 
deforms the helix and a chiral soliton lattice (CSL) appears
\cite{Kishine15,Dzyal58,Dzyal64,Izyumov84,Kishine05}. This CSL, which is realized \cite{Togawa12} in 
\CrNbS, supports phenomena very interesting for spintronics, like spin motive forces \cite{Ovchinnikov13} 
and tuneable magnetoresistence \cite{Kishine11,Togawa13,Ghimire13,Bornstein15}. 
A theoretical analysis at zero temperature carried out long ago \cite{Dzyal64} concluded that
by increasing the field the period of the CSL increases and, eventually, as the period diverges, 
a phase transition takes place continuously to a forced FM (FFM) state. 
This prediction has been recently confirmed experimentally \cite{Togawa12,Togawa13} and
theoretically by computations at finite temperature \cite{Laliena16b}.

If, on the other hand, the applied field is parallel to the DM axis, the local magnetic moment
acquires a constant component along the axis and a conical helix is formed. The conical helix
propagates with a period $L_0$ which is independent of temperature and field intensity.
As the field increases, the component of the local magnetic moment parallel to the DM axis 
increases and the perpendicular component decreases. A transition to a homogeneous FFM state
takes place when the perpendicular component vanishes. The same happens if the temperature
is increased at constant parallel field.

A classification of the continuous transitions that take place between spatially homogeneous and modulated 
states was introduced long ago by DeGennes \cite{DeGennes75}, who called \textit{nucleation} transitions
those in which the period of the modulated state diverges when the transition point is approached
from the modulated phase, and \textit{instability} transitions those in which the phase transition
takes place when the intensity of the 
Fourier modes with non-zero wave vector tend to zero, while the fundamental wavevector 
remains finite and does not vanish. 
The mechanisms for these two kind of transitions are qualitatively
different. In the monoaxial helimagnet, the transition between the CSL and the FFM states  
as a perpendicular magnetic field increases at sufficiently low temperature is of \textit{nucleation}
type \cite{Dzyal64, Laliena16b}. On the other hand, at zero field mean field theory predicts 
an \textit{instability} type continuous transition at the ordering temperature $T_0$.
The transition to the FFM in presence of a parallel magnetic field is also of second order 
\textit{instability} type phase transition.
 
Hence, by varying the temperature from $0$ to $T_0$ and/or the applied field from completely
perpendicular to completely parallel, the transition changes 
from \textit{nucleation} to \textit{instability} type. 
How this change of regime takes place is a very interesting question which may also have
interesting phenomenological consequences.

Recently, the zero temperature phase diagram of the monoaxial helimagnet has been theoretically 
analized for oblique magnetic fields \cite{Laliena16a}, which are neither perpendicular nor parallel
to the DM axis. It has been found that in the thermodynamic space formed by the parallel and perpendicular 
components of the magnetic field two separated continuous transition lines appear. 
The transitions along the line that contains as limiting case the parallel field 
are of \textit{instability} type and the transitions along the other line, 
which contains as limiting case the perpendicular field, are of
\textit{nucleation} type. The two continuous lines are separated by a line of discontinuous transitions.
Two tricritical points separate the discontinuous transition line from the continuous transition lines.

Also recently the phase diagram of the monoaxial helimagnet in the thermodynamic space defined by
the temperature and a perpendicular 
magnetic field  has been theoretically studied in Ref.~\onlinecite{Laliena16b}.
The conclusion is that at low $T$ the transitions to the FFM state induced by 
the perpendicular field are continuos, of \textit{nucleation} type, with the period of the chiral structure 
diverging at the transition points. As temperature increases the critical field decreases and vanishes
at the zero field critical temperature, $T_0$. The transition at $T_0$ is continuous of \textit{instability} 
type.
The transition line in the vicinity of $T_0$ is of first order and it is separated from the low $T$
continuous transition line by a tricritical point. This somehow unexpected behavior is rather logical
as it is difficult to imagine how to connect continuously \textit{instability} and \textit{nucleation} 
transitions.
The prediction of a first order transition and a tricritical point in the vicinity of $T_0$ may be a
clue to the interpretation of the experimental results on the phase diagram reported in 
Refs.~\onlinecite{Tsuruta16,Bornstein15,Ghimire13}.
A more refined numerical computation around the $T_0$ neighborhood, carried out
in this work, leads to the conclusion that the first order line
does not actually end at $T_0$, but instead the transition is of second order 
\textit{instability type} in a very short line that ends at $T_0$. Correspondingly,
a second tricritical point appears separating this short second order line from the first order line.
This tricritical point was unnoticed in Ref.~\onlinecite{Laliena16b}.

In this work we complete the theoretical study of the phase diagram of the monoaxial helimagnet
and the nature of its phase boundaries by analyzing it in the 3D thermodynamic space $H_x-H_z-T$,
where $H_x$ and $H_z$ stand repectively for the perpendicular and parallel components of the magnetic field.
The thermal fluctuations are treated classically and therefore the results are not valid at very low T,
where it is well known that a quantum treatment of thermal fluctuations is necessary, for instance, to
reproduce the behavior of the specific heat. In the zero temperature limit, however, thermal fluctuations 
disappear and the semiclassical approximation seems to describe well the ground state of these kind of 
systems.

The methods presented in this work can be applied to other systems in which phase transitions from
spatially modulated phases to homegeneous phases take place, as for instance cholesteric liquid crystals.

\section{Model\label{sec:model}}

Let us consider the model described in Ref~\onlinecite{Laliena16b}:
a classical spin system with FM exchange and monoaxial DM interactions, 
and single-ion easy-plane anisotropy, at temperature $T$ and in presence of an applied magnetic 
field $\vec{H}$. In what follows we use the notation
of Ref.~\onlinecite{Laliena16b}, and take the $\hat{z}$ coordinate axis along the DM axis.

To get the thermodynamical properties we evaluate the free energy, $\mathcal{F}$, through
the variational mean field approximation, which has been succesfully applied to
the study of the double-exchange model of itinerant ferromagnetism
\cite{Laliena01,laliena01b,Laliena02} and, in combination with
\textit{ab-initio} techniques, to the study of the temperature dependence of thermodynamic quantities
in itinerant ferromagnets\cite{Gyorffy85,Staunton86,Staunton06}.
The free energy is obtained by minimizing the mean field free energy, $\mathcal{F}_0$, with respect
to the mean field configuration, $\vec{M}_{\vec{r}}$.
In the continuum limit, taken along the lines described in Ref.~\onlinecite{Laliena16b},
we get $\mathcal{F}_0=\epsilon_0\int d^3r f_0(\vec{r})$, with
\begin{equation}
f_0=\frac{1}{2}\sum_i\xi_i(\partial_i\vec{m})^2
-q_0\hat{z}\cdot(\vec{m}\times\partial_z\vec{m})
-q_0^2(\vec{h}\cdot\vec{m}+U),
\label{eq:f0}
\end{equation}
where 
\begin{equation}
U=\frac{\mu^2}{2}m^2 - \gamma\left[F+(1-3F)\frac{m_z^2}{m^2}\right]
 + \alpha\left[\ln\frac{\sinh M}{M} - M m\right],
\label{eq:U}
\end{equation}
where $\vec{m}=F\vec{M}$ is the mean local magnetic moment and $F=\coth(M)/M-1/M^2$,

The relation of the parameters entering Eqs.~(\ref{eq:f0}) and~(\ref{eq:U}) with a more
fundamental Hamiltonian is given in Ref.~\onlinecite{Laliena16b}.
The $\xi_i$ measure the spatial anisotropy of the FM exchange couplings.
By definition, $\xi_z=1$, and in this work we consider only system of symmetry such that
$\xi_x=\xi_y=\xi$. The parameter $q_0$ has dimensions of inverse length and gives the propagation
vector of the helical modulation at zero field.
The remaining parameters are dimensionless: 
$\mu^2$ controls the continuum limit and has to be large \cite{Laliena16b}; 
and $\gamma$, $\alpha$, and $\vec{h}$
are proportional to the single-ion anisotropy, temperature ($T$) 
and external magnetic field ($\vec{H}$), respectively.
Finally, $\epsilon_0$ is an overall constant with the dimensions
of energy per unit length.
All these parameters might be obtained from \textit{ab-initio} calculations, but in practice
can be fit to experimental results to describe the phase diagram of different samples
and materials. 

\section{Method of solution \label{sec:method}}

The minimum of $\mathcal{F}_0$ is a solution of the corresponding Euler-Lagrange equations.
Clearly, the mean field configuration which minimizes $\mathcal{F}_0$ depends only on $z$ and the 
equations read
\begin{equation}
\vec{M}^{\prime\prime} = \Omega\vec{M}^\prime + 2q_0\hat{z}\times\vec{M}^\prime + \Psi\vec{M}
+ \Upsilon\hat{z}\times\vec{M} + \Pi M_z\hat{z} - q_0^2\vec{h}/F.
\label{eq:EL}
\end{equation}
The scalar functions $\Omega$, $\Psi$, $\Upsilon$, and $\Pi$ depend on $\vec{M}$ and $\vec{M}^\prime$.
They are given in appendix~\ref{app:EL}.

Equations~(\ref{eq:EL}) constitute a system of three second order differential equations, the general 
solution of which contains six arbitrary integration constants. The task is to find the particular solution 
which minimises $\mathcal{F}_0$. We follow the method described in Ref.~\onlinecite{Laliena16a}.
On physical grounds, we expect a periodic ground state\footnote{Throughout this work we use the term 
\textit{ground state} for the spin configuration which minimizes the mean field free energy.
Although an abuse of language, it is not uncommon to use the term ground state at finite temperature
to refer to the set of equilibrium correlation functions determined by the density matrix.}, 
with period $L$. 
Hence, the free energy density $\bar{f}_0=\mathcal{F}_0/V$, where $V$ is the volume, is equal to the 
free energy averaged over one period, that is $\bar{f}_0=(1/L)\int_0^Lf_0(z)dz$, and the boundary conditions 
(BC) are $\vec{M}(0)=\vec{M}(L)$. Since the equations are of second order, 
these BC do not guarantee periodicity, which requires also the equality of the first
derivatives at the two boundaries: $\vec{M}^\prime(0)=\vec{M}^\prime(L)$. 
These additional conditions cannot be generally imposed on the boundary value problem (BVP), since it would 
be overdetermined. 
The strategy to find a solution to the problem is as follows: with no loss, set $M_y(0)=M_y(L)=0$,
and for given $L$ and $M_x(0)=M_{x0}$, and $M_z(0)=M_{z0}$, solve numerically the BVP; for fixed $L$, 
tune $M_{x0}$ and $M_{z0}$ until periodicity is reached; then, compute $\bar{f}_0$ via a numerical 
quadrature algorithm. 
The equilibrium period is the minimum of $\bar{f}_0$, which is found via a simple 
minimisation algorithm. 

\section{Phase diagram \label{sec:phd}}

Without any loss, we can choose a magnetic field with components along $\hat{x}$ and $\hat{z}$, and
set $h_y=0$. Two phases appear: the homogeneous FFM state at high temperature and/or high field
and a spatially modulated structure at low temperature and low field, which
is generically named here helicoid.

A surface of phase transitions in the thermodynamic space $(T,H_x,H_z)$
separates the helicoid and FFM phases.
The transition surface can be described by giving one of the thermodynamic coordinates as a 
function of the other two. The dependent coordinate is denoted by $T_c$, or $H_{xc}$, or $H_{zc}$. 


The transition points on the three axes of the thermodynamic space can be analytically obtained:
the critical temperature at zero field, $\alpha_0$, and the critical perpendicular and parallel fields at
zero temperature, $h_{x0}$ and $h_{z0}$, respectively.
Their analytic expressions are:
\begin{equation}
\alpha_0 = \frac{1}{3}(\mu^2 + 1) + \frac{2}{15}\gamma, \quad
h_{x0} = \frac{\pi^2}{16}, \quad
h_{z0} = 1 + 2\gamma. 
\label{eq:params}
\end{equation}
The corresponding dimensionfull quantities, denoted by $T_0$, $H_{x0}$, and $H_{z0}$, 
are directly measurable quantities: zero field critical temperature and zero
temperature critical perpendicular and parallel fields, respectively.
It is convenient to present the results in terms of $T/T_0=\alpha/\alpha_0$,
$H_x/H_{x0}=h_x/h_{x0}$, and $H_z/H_{z0}=h_z/h_{z0}$.
We also use the notation $H_0=\sqrt{H_{x0}^2+H_{z0}^2}$, and denote by $\theta$ the angle formed
by the magnetic field and the DM axis: 
$\tan\theta=H_x/H_z$. 
Except in the $H_z=0$ separate discussion given below, all the results presented here correspond 
to $\mu^2=210$, which is a value appropriate to describe the phenomenology of 
\CrNbS \cite{Laliena16b}.

\begin{figure}[t!]
\centering
\includegraphics[width=\linewidth,angle=0]{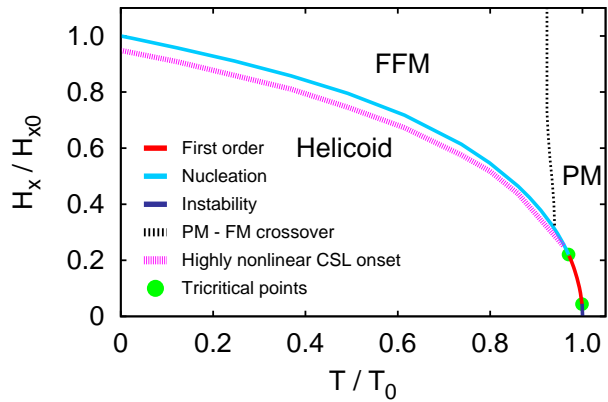}
\caption{$H_x-T$ phase diagram for $H_z=0$ (perpendicular field) calculated with $\mu^2=120$ 
and $\gamma=2.58$. 
The light and dark blue lines correspond to second order \textit{nucleation} and 
\textit{instability} type transitions, respectively,
and the red line to first order transitions.
The pink dotted line marks the onset of the highly nonlinear CSL.
The dashed black line signals the crossover from PM to FM behavior.
The tricritical point closer to $T_0$ was unnoticed in 
Ref.~\onlinecite{Laliena16b}.
\label{fig:phdhx}}
\end{figure}


For $H_z=0$ we have re-analyzed the transition line in the very close neighborhood 
of the zero field transition
with more detail and accuracy than in Ref.~\onlinecite{Laliena16b}. In this region there is
instability caused by the critical fluctuations and the numerical computations are more difficult.
It turns out that
a second tricritical point, not detected in Ref.~\onlinecite{Laliena16b}, appears at $T/T_0\approx 0.9989$ 
and $H_x/H_{x0}\approx 0.027$. Hence, the transition line is of second order nucleation type
at low temperature and of second order instability type at high temperature. These two second order lines
are separated by a first order line, and two tricritical points separate the first order line from
the second order lines.
The phase diagram is displayed in Fig.~\ref{fig:phdhx}.
Notice the slight difference with the phase diagram published in Ref.~\onlinecite{Laliena16b}.
Although the region around the zero field phase transition, in which the fluctuations are expected
to be strongly correlated, is probably not well described by mean field theory, these results give 
a hint on what can be expected, before more sophisticated approaches,
like Monte Carlo simulations, are fully developed. Work in this direction has been reported in
Ref.~\onlinecite{Nishikawa16}.

\begin{figure}[t!]
\centering
\includegraphics[width=\linewidth,angle=0]{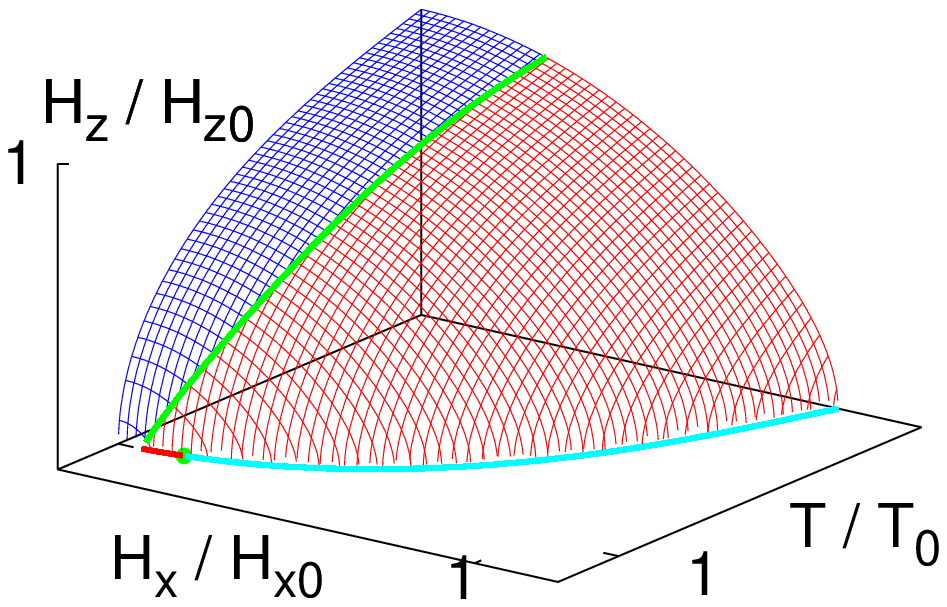}
\includegraphics[width=\linewidth,angle=0]{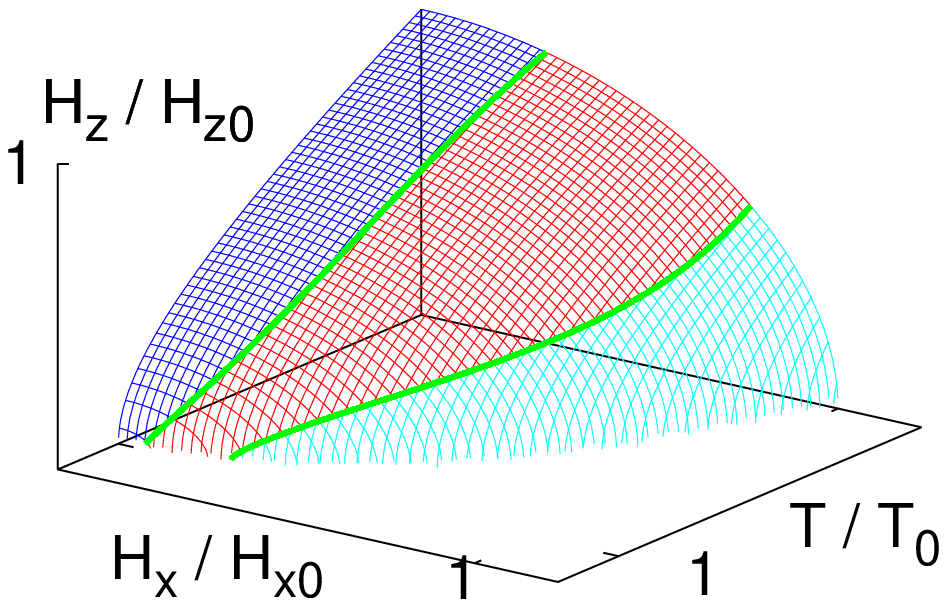}
\caption{3D phase diagram without (top) and with (bottom) single-ion anisotropy ($\gamma=2.58$). 
The second order transitions take place on the
dark blue (\textit{instability}) and light blue (\textit{nucleation}) portions of the transition surface. 
On the red portion the transitions are of first order.
The tricritical lines separating the first order surface from the two second order surfaces
are displayed in green.
\label{fig:phd3D}}
\end{figure}

For $H_x=0$ the free energy is minimized by a conical helix with pitch $L_0$
independent of $\alpha$ and $h_z$. The angle $\theta_0$ which forms $\vec{m}$ 
with the DM axis depends on temperature and magnetic field.
The transition to the FFM state takes place continuously as $\cos\theta_0\rightarrow 1$ and
is of \textit{instability} type. An order parameter which vanishes in the FFM state is 
$\sin\theta_0$. 
As the transition line is approached $\sin\theta_0$ vanishes as a power law, 
with the mean field exponent:  
$\sin\theta_0\sim (1-\alpha/\alpha_c)^{1/2}$ or
$\sin\theta_0\sim (1-h_z/h_{zc})^{1/2}$.

Three dimensional representations of the phase diagram without ($\gamma=0$) and with ($\gamma=2.58$) 
anisotropy are displayed on the top and botom panels of Fig.~\ref{fig:phd3D}, respectively.
The value $\gamma=2.58$ has been chosen so that the critical parallel field 
of \CrNbS\ at relatively low $T$ is reproduced \cite{Laliena16a}.

Let us discuss first the general case with non vanishing anisotropy.
The transition surface is divided into three parts: 
two surfaces of second order transitions are separated by a surface of first order transitions. 
The second order transitions of the surface
that intersects the $H_x=0$ plane are of \textit{instability} type, while the transitions 
of the other second order surface are of \textit{nucleation} type. 
The \textit{instability} surface is separated from the 
first order surface by a line of tricritical points that is called the \textit{instability} tricritical 
line (ITC).
Analogously, the boundary between the \textit{nucleation} and the first order surfaces
is a line of tricritical points called the \textit{nucleation} tricritical line (NTC).
The tricritical points TC$_\mathrm{I}$ and TC$_\mathrm{N}$ found at $T=0$ in Ref.~\onlinecite{Laliena16a}
belong to ITC and NTC, respectively.

As $\gamma\rightarrow 0$ the \textit{nucleation} surface shrinks and in the absence of 
single-ion anisotropy (Fig.~\ref{fig:phd3D}, top) is squeezed onto a line on the $H_z=0$ plane.
The transition surface contains a second order \textit{instability} surface and a first order 
surface separated by the ITC line. The NTC line is reduced to a point on the
$H_z=0$ plane.

In terms of $T/T_0$, $H_x/H_{x0}$, and $H_z/H_{z0}$, the shape of the transition surface is 
nearly independent of $\mu^2$ provided that $\mu^2$ is large enough.
It depends, however, on the value of the single-ion anisotropy, although this dependence
disappears gradually as the anisotropy grows; in this case it shows noticeably dependence on $\gamma$
only for $T$ close to $T_0$. This dependence is related to the fluctuations of 
$m_z^2$ at high $T$.

The structure of the modulated state depends on temperature and magnetic field. 
For fields with small perpendicular component, 
it is a slightly distorted conical helix, a quasilinear structure to which only a few 
Fourier harmonics give a noticeably contribution. As the perpendicular 
component is gradually increased, higher order Fourier harmonics appear and
the helix becomes a conical CSL. A highly nonlinear
CSL, receiving appreciably contributions from many Fourier harmonics, appears
only in the vicinity of the \textit{nucleation} surface, in complete similarity with the $H_z=0$ 
case \cite{Laliena16b}.
The highly nonlinear CSL regime is not sharply defined, but separated from the rest of the 
modulated phase by a crossover surface very close to the \textit{nucleation} surface.
This crossover surface is not shown in Fig.~\ref{fig:phd3D}, but its intersection with the 
$H_z=0$ plane is shown in Fig.~\ref{fig:phdhx} (highly nonlinear CSL onset line).




\section{Singularities on the transition surface}

\begin{figure}[t!]
\centering
\includegraphics[width=\linewidth,angle=0]{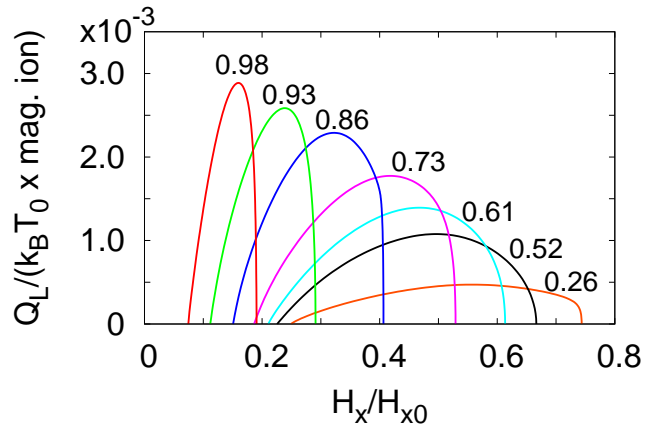}%
\caption{Latent heat per magnetic ion, in units of $k_\mathrm{B} T_0$, along the first order 
isothermal transition lines for $T/T_0$ indicated in the plot. 
Parameters: $\mu^2=210$ and $\gamma=2.58$.
\label{fig:latHeat}}%
\end{figure}

\begin{figure}[t!]
\centering
\includegraphics[width=\linewidth,angle=0]{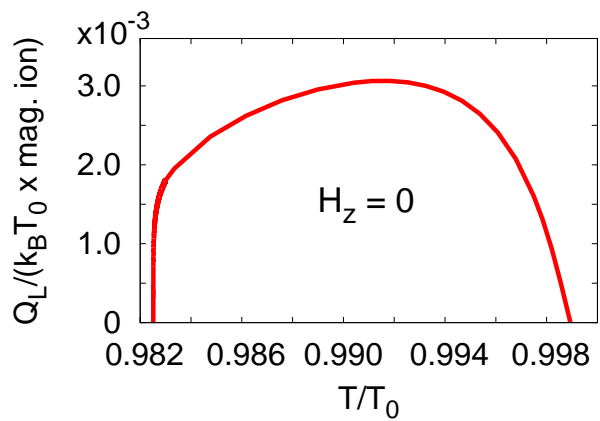}%
\caption{Latent heat per magnetic ion, in units of $k_\mathrm{B} T_0$, as a function of temperature
along the first order transition line for perpendicular field ($H_z=0$).
It vanishes before reaching $T_0$, at a tricritical point 
in the close vicinity of $T_0$, not detected
in Ref.~\onlinecite{Laliena16b}. Parameters: $\mu^2=210$ and $\gamma=2.58$. 
\label{fig:latHeat_Hz0}}%
\end{figure}

On the first order transition surface the helicoid and FFM states coexist.
On both sides of the transition surface the two states are present, one as stable and the other
as metastable state.
As a consequence of the different entropies of these two states, a latent heat accompanies the 
transition. The latent heat vanishes on the boundaries of the first order surface 
(the tricritical lines) and thus reaches a maximum at an interior point 
of each isothermal transition line, as can be seen in Fig.~\ref{fig:latHeat}. 
By increasing $T$ the latent heat maximum increases and its position is shifted towards
smaller values of $H_x$ and $H_z$\footnote{Recall that each pair $(T,H_x)$ defines a value of $H_z$ since 
the transition line lies on the transition surface.}. The absolute maximum, reached at $H_z=0$, 
is about $3\times 10^{-3}\,k_\mathrm{B}T_0$ per magnetic ion, what amounts to
6 J/kg in the case of \CrNbS ($T_0\approx 125$ K). 
Fig.~\ref{fig:latHeat_Hz0} displays the latent heat as a function of $T/T_0$ for $H_z=0$.
Notice the slight difference with the analogous figure of Ref.~\onlinecite{Laliena16b},
due to the refined computations in the vicinity of $T_0$.

\begin{figure}[t!]
\centering
\includegraphics[width=\linewidth,angle=0]{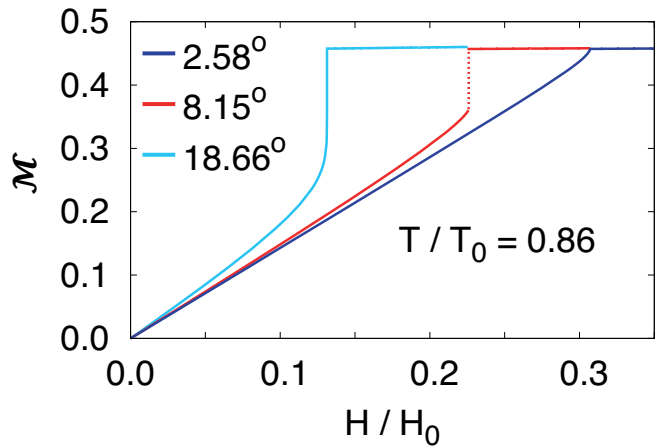}%
\caption{Magnetization per magnetic ion as a function of the field strength for three
directions of the field and $T/T_0=0.86$.
The legend shows the corresponding value of the angle ($\theta$) between the field and the DM axis.
The transitions are of second order \textit{instability} type (dark blue),
of first order (red), and of second order \textit{nucleation} type (light blue).
Parameters: $\mu^2=210$ and $\gamma=2.58$.
\label{fig:magnet_H}}%
\end{figure}

\begin{figure}[t!]
\centering
\includegraphics[width=\linewidth,angle=0]{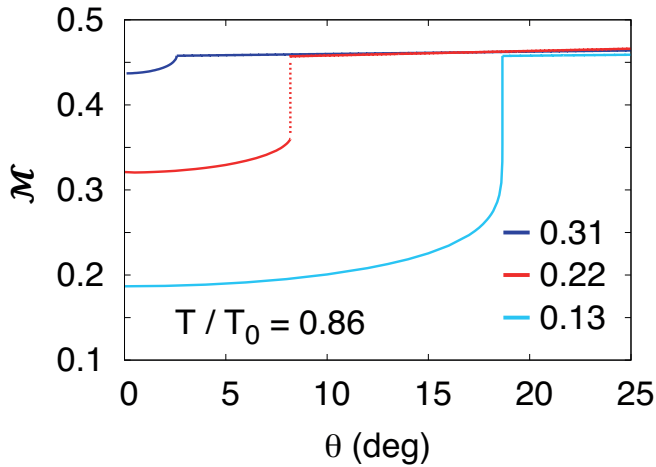}%
\caption{Magnetization per magnetic ion as a function of the angle ($\theta$) between the
field and the DM axis, for three field strengths and $T/T_0=0.86$.
The legend shows the corresponding values of $H/H_0$.
The transitions are of second order \textit{instability} type (dark blue),
of first order (red), and of second order \textit{nucleation} type (light blue).
Parameters: $\mu^2=210$, $\gamma=2.58$.
\label{fig:magnet_angle}}%
\end{figure}

Fig.~\ref{fig:magnet_H} displays the behavior of the magnetization per magnetic ion, 
$\mathcal{M}=(1/L)|\int_0^L\vec{m}(z) dz|$, 
as a function 
of the field strength 
for three fixed field directions, corresponding to phase 
transitions of \textit{instability} type (dark blue), of first order (red), and of
\textit{nucleation} type (light blue).
Fig.~\ref{fig:magnet_angle} shows $\mathcal{M}$ versus the field direction at constant field strength, 
for three values of the field strength that are representative of \textit{instability} (dark blue), 
first order (red), and \textit{nucleation} (light blue) phase transitions. 
The field direction is characterized by the angle $\theta$ that forms with the DM axis.

On the transition surface $\mathcal{M}$ is singular. It presents a finite discontinuity,
signaled by the broken line in Figs.~\ref{fig:magnet_H} and~\ref{fig:magnet_angle},
on the first order surface. On the two second order surfaces $\mathcal{M}$ is continuous, but
attains the value of the FFM magnetization in a singular way. 
On the \textit{instability} surface the singularity is not very sharp, and, numerically, it seems 
to be described by a power law, with a critical exponent between 1/2 and 2/3.

The singularity on the \textit{nucleation} surface is controlled by the divergence 
of the period, $L$, since the difference between the CSL and FFM magnetization scales as $1/L$.
An analysis of the numerical results shows that when the transition point is approached 
by tuning a parameter $\zeta$, the period $L$ satisfies the scaling law 
\begin{equation}
B(Aq_0L+1)\exp(-Aq_0L)\sim (\zeta_c-\zeta)/\zeta_c, \label{eq:scaling}
\end{equation}
where $\zeta$ can be either $T$, or $H_x$, or $H_z$, or $H$, or $\theta$.
It is motivated by the well known logarithmic singularity \cite{Dzyal64}
that appears as $H_x\rightarrow H_{x0}$ at $T=0$ and $H_z=0$. \cite{Laliena16a,Laliena16b}
The scaling of $L$ (and therefore of $\mathcal{M}$) is thus a universal feature of the CSL.
It is interesting that the coefficient $A$ depends only on the transition point, 
and not on the parameter $\zeta$ tuned to reach it. 

The inverse of the period, $1/L$, is the density of solitons. It was shown in 
Ref.~\onlinecite{Laliena16b} that for a purely perpendicular field the density of solitons 
is a universal function of $H_x/H_{xc}$, independent of $T$, for temperatures below the 
nucleation tricritical temperature. 
Above this tricritical temperature, universality is lost. 
This universality also holds when the field has a component along 
the DM axis, provided that the transition point reached by increasing $H_x$ while $T$ and 
$H_z$ are kept constant lies on the \textit{nucleation} surface.
It is obvious that the universality cannot hold in the whole phase diagram since in the 
vicinity of the \textit{instability} surface $L$ is almost independent of the field. 
Therefore, the lost of universality is a way of locating the
\textit{nucleation} tricritical line.

The universality of the magnetoresistance curves of \CrNbS\ in presence of a perpendicular
field reported in Ref.~\onlinecite{Togawa13} was linked to the universality 
of the soliton density curves. Measurements of the magnetorresistance with oblique fields
can be used to verify experimentally the universality predicted in the present paper and to 
locate the \textit{nucleation} tricritical line of \CrNbS.

\section{Specific heat}

The specific heat can be computed as $C_\mathrm{V}=T\partial s/\partial T$, where $s$ is the 
specific entropy (per unit mass), which, in the mean field approach,
is given by
\begin{equation}
s = \frac{k_\mathrm{B}}{\rho v} \frac{1}{L}\int_0^L \left[\ln(\sinh M/M) - FM^2\right] dz,
\label{eq:entropy}
\end{equation}
where $\rho$ is the mass density and $v$ the volume of the unit cell of the underlying lattice
and $\vec{M}(z)$ is the equilibrium mean field configuration (\textit{i.e.}, the solution of the 
Euler-Lagrange equations that minimizes the free energy).
This configuration depends in principle on three parameters, the period $L$ and the two BCs,
$M_{x0}$ and $M_{z0}$.
However, as discussed in section~\ref{sec:method}, only one of these three parameters
can be independently chosen. For a given value of $L$, the two BCs
are determined by the requirement of periodicity. 
Hence, $\vec{M}$ is a function of $z$ and $L$, which in its turn is a function of the temperature 
and the field. As a result, all the $T$ dependence of $s$ comes from its $L$ dependence. 
Hence we have
\begin{equation}
C_\mathrm{V} = \frac{\partial s}{\partial L} T \frac{\partial L}{\partial T} 
\end{equation}
On the \textit{nucleation} surface $T\partial L/\partial T$ diverges.
From Eq.~(\ref{eq:scaling}), with $\zeta=T$, we get that for $T\rightarrow T_c$
\begin{equation}
T\frac{\partial L}{\partial T} \sim \frac{1}{A} \frac{T_c}{T_c-T}.
\end{equation}
An expression for the factor $\partial s/\partial L$ can be readily obtained 
from~(\ref{eq:entropy}):
\begin{eqnarray}
\frac{\partial s}{\partial L} &=& \frac{k_\mathrm{B}}{\rho v}
\frac{1}{L}\int_0^L GM\frac{\partial M}{\partial L} dz \nonumber \\
&+& \frac{1}{L} \left\{ \frac{k_\mathrm{B}}{\rho v}\left[\ln(\sinh M_0/M_0)-FM_0^2\right]-s
\right\}, \label{eq:partialL}
\end{eqnarray}
where $M_0=\sqrt{M_{x0}^2+M_{z0}^2}$ is the value of $M$ at the boundaries
$z=0$ and $z=L$, and
we used the fact that the derivative with respect to $M$ of the integrand of 
Eq.~(\ref{eq:entropy}) is $GM$, with $G=F+MdF/dM$.



\begin{figure}[t!]
\centering
\includegraphics[width=\linewidth,angle=0]{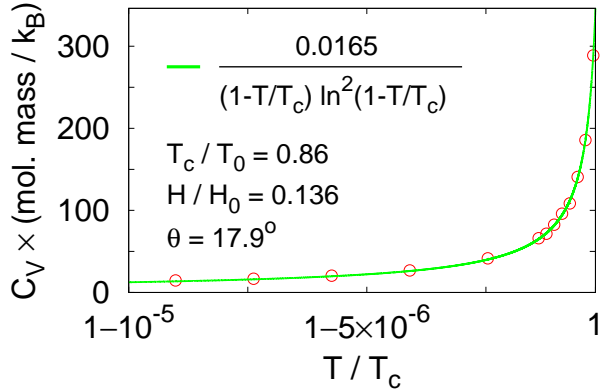}
\caption{
Divergence of the specific heat as a \textit{nucleation} transition point is approached.
The legend displays the value of the field, which is kept fixed, and of the transition temperature.
Parameters: $\mu^2=210$ and $\gamma=2.58$.
\label{fig:scaling_calor}}
\end{figure}

In the $T\rightarrow T_c$ limit $M$ and $M_0$ tend to the FFM mean field as $1/L$.
Thus, the two terms of the right hand side of Eq.~(\ref{eq:partialL}) vanish as $1/L^2$
(the term in curly braces tends to the difference of the CSL and FFM specific entropies
and thus vanishes as $1/L$). This simply means that, since $s$ tends to the FFM entropy
as $1/L$, its derivative with respect to $L$ vanishes as $1/L^2$. 
Thus, taking into account that $L\sim\ln(T_c-T)$, the specific heat diverges on the 
\textit{nucleation} surface as 
\begin{equation}
C_\mathrm{V}\sim 1/[(T_c-T)\ln^2(T_c-T)].
\end{equation}
The numerical computations confirm this behavior. A fit of the parameters 
$c_0$ and $b$ of the function
\begin{equation}
c_0 / [(1-T/T_c)\ln^b(1-T/T_c)]
\end{equation}
to the computed $C_\mathrm{V}$ for fixed field in the region close to $T_c$ gives 
$b=2.08\pm0.078$, which is perfectly compatible with $b=2$. Therefore, we may fix $b=2$ and 
fit the single parameter $c_0$. The result is displayed in Fig~\ref{fig:scaling_calor}. 
We observe a perfect agreement of the numerical results with the theoretical expectation.
The divergence of the specific heat found here is remarkable since in the ``canonical'' mean field 
theory of the PM-FM transition the specific heat has no divergence, but shows a finite 
discontinuity at the critical point. 

On the first order surface the specific heat has obviously
a finite discontinuity. It shows also a finite discontinuity on the \textit{instability} surface,
since $L$ and its derivatives remain finite there. For zero field, the specific heat jump at $T=T_0$
can be analytically computed as follows.
The low $T$ ground state is an helix with pitch $L_0$ and
$M(z)=M_0$ independent of $z$, determined by
\begin{equation} 
(\mu^2+1)F(M_0) - \gamma F_1(M_0)/[M_0G(M_0)] = \alpha. \label{eq:zerofield}
\end{equation}
The left hand side of the above equation attains its maximum value at $M_0=0$, which gives
$\alpha_0$ defined in Eq.~(\ref{eq:params}).
Thus, for $\alpha>\alpha_0$ Eq.~(\ref{eq:zerofield})
has no solution and the system is in the PM phase.
The solution of~(\ref{eq:zerofield}) decreases monotonically with $\alpha$ 
from $M_0=\infty$ (what implies saturation of magnetization, $m=1$) at $\alpha=0$
to $M_0=0$ at the critical point $\alpha_0$. 
The transition to the PM phase at $\alpha_0$ takes place continuously and
$M_0$ vanishes as a power law: $M_0\sim(1-T/T_0)^{1/2}$.
It is a second order \textit{instability}
type transition. The specific entropy is given by
\begin{equation}
s = \frac{k_\mathrm{B}}{\rho v} \left[\ln(\sinh M_0/M_0) - FM_0^2\right],
\end{equation} 
and the specific heat by
\begin{equation}
C_\mathrm{V} = \frac{k_\mathrm{B}}{\rho v} G M_0\alpha\frac{\partial M_0}{\partial\alpha}.
\end{equation}
Implicit differenciation of Eq.~(\ref{eq:zerofield}) with respect to $\alpha$ gives
\begin{equation}
\frac{\partial M_0}{\partial\alpha} = 
\frac{G^2M_0^2/F_1}{1+\mu^2+\gamma\left(F+3M_0F_1+M_0^2F_2-GF_2M_0/F_1\right)}.
\end{equation}
For $T\rightarrow T_0$ (\textit{i.e.} $\alpha\rightarrow\alpha_0$) we have 
$M_0\rightarrow 0$, and to leading order in $1/M_0$ the above equation gives
\begin{equation}
\frac{\partial M_0}{\partial\alpha} \sim \frac{45/2}{1+\mu^2+(2/35)\gamma}\frac{1}{M_0},
\end{equation}
so that
\begin{equation}
C_\mathrm{V} = \frac{k_\mathrm{B}}{\rho v} \frac{5}{2} 
\left[\frac{\mu^2+1+(2/5)\gamma}{\mu^2+1+(2/35)\gamma}+O(T_0-T)\right], \label{eq:shh0}
\end{equation}
where we have substituted $G$ by 1/3, which is its value at $M=0$, and $\alpha$ by
the value $\alpha_0$ given by Eq.~(\ref{eq:params}). In the PM phase mean field theory gives
$M_0=0$, $s=0$, and $C_\mathrm{V}=0$, and therefore the specific heat jump at the zero
field critical point is given by Eq.~(\ref{eq:shh0}).
Since $\mu^2\gg 1$, the specific heat jump is nearly independent of $\mu^2$ and
$\gamma$, and is given by $\Delta C_\mathrm{V} \approx (5/2) k_\mathrm{B}/\rho v$. 

\begin{figure}[t!]
\centering
\includegraphics[width=\linewidth,angle=0]{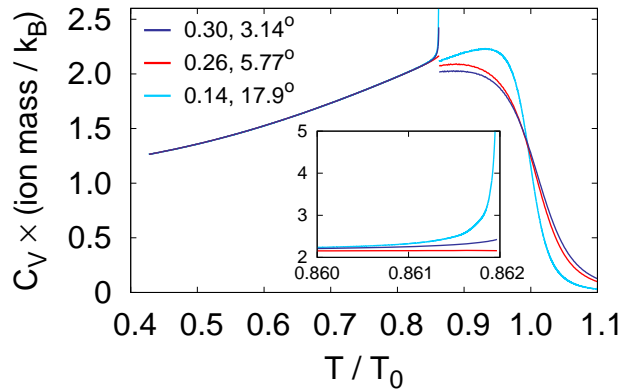}
\caption{
Specific heat \textit{vs.} $T/T_0$ for three fixed values of the field 
specified in the legend, in polar coordinates ($H/H_0$, $\theta$). 
In the three cases the phase transition takes place at $T_c/T_0=0.862$ and are of
second order \textit{instability} type (dark blue), of first order (red),
and of second order \textit{nucleation} type (light blue). 
The inset displays the behavior in the vicinity of the transition.
Parameters: $\mu^2=210$ and $\gamma=2.58$.
\label{fig:calor}}
\end{figure}

The behavior of the specific heat as a function of $T/T_0$ for fixed field is 
displayed in Fig.~\ref{fig:calor} for the three values of the field shown in the
legend. In the three cases the transition temperature is the same, $T_c/T_0=0.862$.
The phase transitions are of \textit{instability} type (dark blue), of first order
(red), and of \textit{nucleation} type (light blue). 
In the helicoid phase the specific heat is basically independent of the field, except 
in the close vicinity of the phase transition, where it shows a rapid growth
in the case of the second order transitions of both types.
However, as can be appreciated in the inset of Fig.~\ref{fig:calor}, 
which shows the behavior of $C_\mathrm{V}$ around $T_C$, the specific heat diverges 
in the case of the \textit{nucleation} transition but remains finite in the 
\textit{instability} case. 

In the low field case (light blue line), the specific heat presents
a broad shoulder in the high temperature phase that is associated to the 
crossover from PM to FFM behavior. This defines a crossover surface in the 3D 
phase diagram, which is not shown in Fig.~\ref{fig:phd3D}. Its intersetion with the
$H_z=0$ plane, however, is shown in Fig.~\ref{fig:phdhx}.

\section{Summary and conclusion \label{sec:conc}}

A complete characterization of the phase diagram of the monoaxial helimagnet 
in the presence of a magnetic field with components parallel and perpendicular to the 
DM axis has been obtained by means of the variational mean field approach. 
The phase diagram contains a low-field and low-temperature phase in which the 
ground state is a spatially modulated chiral magnetic structure and a 
high-field/high-T phase in which the system is in a homogeneous forced ferromagnetic state
(paramagnetic at zero field and high temperature). The phase boundary is a surface in the
three dimensional thermodynamic space defined by the temperature and the parallel and 
perpendicular components of the magnetic field. The transition surface is divided into
three parts: one surface of first order transitions separates 
two surfaces of second order transitions, in one of which the transitions are of 
\textit{instability} type and in the other one of \textit{nucleation} type.
The first order surface is separated from the second order surfaces by two lines of 
tricritical points. 

It is worthwhile to recall that mean field theory, which approximates the thermal fluctuations 
by the uncorrelated fluctuations of the trial ``Hamiltonian'', usually fails in the critical domain,
where the fluctuations are strongly correlated.
In our case, except for a small neighborhood of the zero field transition,
the fluctuations are not expected to be critical, since the transitions 
are driven by the magnetic field,
and the computations should be accurate, or at least qualitatively correct.
Only in the vicinity of the zero field phase transition, where critical fluctuations are expected,
mean field theory may fail, and other techniques, as Monte Carlo simulations, 
are necessary to validate or disproof the mean field results.

The period of the modulated state diverges on the \textit{nucleation} surface.
The divergence obeys a logarithmic scaling law, Eq.~(\ref{eq:scaling}),
which is a distinct feature of the CSL. It induces a singularity in the 
magnetization also characteristic of the CSL.
The specific heat is also divergent on the \textit{nucleation} surface, with a scaling law that
has the form of a power law with logarithmic corrections.
This is remarkable since in the ``canonical'' mean field 
theory of the FM-PM transition the specific heat does not diverge, but shows a finite 
discontinuity. 

The soliton density is a universal function of the reduced perpendicular component of 
the field, $H_x/H_{xc}$. Universality means independence of the temperature and of the 
parallel component of the field, and holds only if the transition point obtained by
tuning $H_x$, keeping fixed $T$ and $H_z$, lies on the 
\textit{nucleation} surface. Otherwise, universality is lost.
Thus, this universality can be used to locate the tricritical line that separates the 
first order surface from the \textit{nucleation} surface.

The picture that emerge from this work should serve to stimulate the experimental
study of the magnetic properties of compounds like \CrNbS, and to interpret some
of the already known and forthcoming experimental data. For instance, the phase diagram
in the immediate vicinity of the zero field ordering transition has a complex structure, 
with first and second order transitions separated by a tricritical point,
and deserves a thorough experimental investigation. 

\begin{acknowledgments}
The authors acknowledge the Grant No. MAT2015-68200-C2-2-P from the Spanish
Ministry of Economy and Competitiveness.
This work was partially supported by the scientific JSPS Grant-in-Aid for 
Scientific Research (S) (No. 25220803), 
and the MEXT program for promoting the enhancement of research universities, 
and JSPS Core-to-Core Program, A. Advanced Research Networks.
\end{acknowledgments}

\appendix

\section{Some details about the Euler-Lagrange equations \label{app:EL}}

In this appendix we give some details about the derivation of the Euler-Lagrange equations,
Eq.~(\ref{eq:EL}).
On physical grounds, the minimum of 
$\mathcal{F}_0=\epsilon_0\int d^3r f_0(\vec{r})$, with $f_0$ given by
Eq.~(\ref{eq:f0}), is a function of the coordinate
$z$ along the DM axis only.
Hence, in the search for the minimum
we can restrict the functional to mean field configurations that depend only on
$z$, and the functional reads $\mathcal{F}_0= A \epsilon_0\int dz f_0(z)$, where
$A=\int dxdy$ is the area of the sample cross section perpendicular to the DM axis and
\begin{equation}
f_0(z)=\frac{1}{2}\vec{m}^{\prime\,2}
-q_0\hat{z}\cdot(\vec{m}\times\vec{m}^\prime)
-q_0^2(\vec{h}\cdot\vec{m}+U),
\label{eq:f0z}
\end{equation}
where the prime stands for the derivative with respect to $z$ and $U$ is given by 
Eq.~(\ref{eq:U}).
The Euler-Lagrange equations then read
\begin{equation}
\frac{d}{dz}\frac{\partial f_0}{\partial M_i^\prime} - \frac{\partial f_0}{\partial M_i} = 0
\end{equation}
for $i\in\{x, y, z\}$. Since $\vec{m}=F\vec{M}$, with $F=\coth(M)/M-1/M^2$, we have
\begin{equation}
\vec{m}^{\prime\,2} = F^2 \vec{M}^\prime\cdot \tilde{T}\vec{M}^\prime
+ G^2 \vec{M}^\prime\cdot \tilde{L}\vec{M}^\prime
\end{equation}
where $G=F+MF_1$, with $F_1=dF/dM$, and
the matrices $\tilde{T}$ and $\tilde{L}$ are respectively the orthogonal projectors 
onto the subspaces transverse and longitudinal to $\vec{M}$:
\begin{equation}
\tilde{T}_{ij} = \delta_{ij} - \frac{M_iM_j}{M^2}, \qquad
\tilde{L}_{ij} = \frac{M_iM_j}{M^2}.
\end{equation}
The Euler-Lagrange equations have then the form
\begin{equation}
(F^2 \tilde{T} + G^2 \tilde{L}) \vec{M}^{\prime\prime} = \vec{W}(\vec{M},\vec{M}^\prime),
\end{equation}
where the vector $\vec{W}$ depends on $\vec{M}$ and $\vec{M}^\prime$, but not on $\vec{M}^{\prime\prime}$.
It is obtained in an straightforward way,
but has a lengthy expression and
is not explicitely written here. The matrix entering the left hand side of the above equation can be
readily inverted, and we get an explicit equation for $\vec{M}^{\prime\prime}$:
\begin{equation}
\vec{M}^{\prime\prime} = \left[\frac{1}{F^2} \tilde{T} + \frac{1}{G^2} \tilde{L}\right]
\vec{W}(\vec{M},\vec{M}^\prime),
\end{equation}
which has the form of Eq.~(\ref{eq:EL}). The explicit form of the functions entering
Eq.~(\ref{eq:EL}) are readliy obtained from the above equation.
Defining $F_2=d^2F/dM^2$, they read
\begin{eqnarray}
\Omega &=&-2(F_1/F)M^\prime, \\
\Psi &=& \Phi+q_0^2(\alpha-\mu^2F)+\Theta M_z^2/M^2, \\
\Upsilon &=& 2q_0(F_1/F)M^\prime, \\
\Pi &=&  2 q_0^2\gamma (1-3F)/(F^2M^2).
\end{eqnarray}
where $M^\prime=dM/dz=\vec{M}^\prime\cdot\vec{M}/M$, 
\begin{equation}
\Theta=-\Pi-3 q_0^2\gamma F_1/(MG^2),
\end{equation}
and
\begin{eqnarray}
& &\Phi = \frac{1}{G}\left[-\frac{F_1}{M}(\vec{M}^{\prime2}-M^{\prime2})
+(2F_1^2/F-F_2)M^{\prime2}
\right. \nonumber \\
& &\left. +2q_0\frac{F_1}{M}\,\hat{z}\cdot(\vec{M}\times\vec{M}^\prime) 
+q_0^2\frac{F_1}{FM}\vec{h}\cdot\vec{M}\right] 
\nonumber \\
& &+q_0^2\gamma\frac{F_1/M}{G^2}.
\label{eq:Phi}
\end{eqnarray}

\vfill\eject




\bibliography{references}

\end{document}